\documentclass[12pt]{article}

\usepackage[english]{babel}
\usepackage[intlimits]{amsmath}

\begin{document}

\title{IS IT POSSIBLE TO TRANSFER AN INFORMATION WITH THE VELOCITIES EXCEEDING
 SPEED OF LIGHT IN EMPTY SPACE?}
\author{L.Ya.Kobelev \footnote{E-mail: leonid.kobelev@usu.ru}  \\
Department of  Physics, Urals State University \\ Lenina Ave., 51,
Ekaterinburg 620083, Russia}
\date{}
.
\maketitle

 abstract:On the base of the theory of time and space with the fractional dimensions
a possibility for information transferring with any velocities is
demonstrated.\\

 01.30.Tt.05.45, 64. 60.A; 00.89.98.02.90. + p

\section{Introduction}
In the theory of special relativity (SR) the maximal velocity of any
signal does not exceed speed of light in the empty space (the existence of
optical taxions does not break SR ) In the frame of multifractal theory of
time and space \cite{1} it is possible to construct the theory of almost
inertial systems \cite{2}. In this theory an arbitrary velocities of
moving particles are possible if the approximate independence of speed of
light from the velocity of the light source and the approximate constancy
of speed of light in vacuum are valid (the breaking the low of constancy
speed of light are less than possibility of modern experiment and consist
$\sim10_{-10}c$, see  \cite{1}). Is the transfer of the information within
the framework of the theory \cite{2}-\cite{4} possible with any
velocities? The difficulty of create a signal carrier of the information
spreading with arbitrary large (practically infinitely large) velocity is
not the main difficulty at the answer to this question. These signals can
be, for example, a beams of charged particles (protons, ionized atoms)
accelerated up to velocities greater then the speed of light (their energy
must be more then energy $E_{0}10^{3}$ where $E_{0}$=$m_{o}c^{2}$) and
then spontaneously accelerated at almost infinite quantity velocity. These
beams may be the carriers of the transferring information. The difficulty
consists in the creating the receivers (detectors) of the information
recorded by beams (or single) faster than light particles. According to
the theory \cite{2}-\cite{4} a particle with velocity $v >c$ is
spontaneously accelerated up to the velocity $v=\infty$ and practically
ceases to interact with a surrounding medium. The purpose of this paper is
the attempt to analyze some opportunities of detection of such particles.
If the problem of detectors for registration of the faster than light
particles will be decided, the problem of practically instantaneous
transfer of the information at any distances is solved positively.

\section{ What physical effects are existing for detection of
particles moving with velocity $v>c$?}

     Let us suppose validity of the laws of the electrodynamics  for the
velocities $v>c$. After replacing $\beta = \sqrt {1 - v^{2}/c^2 }$  by
$\beta^{*} = \sqrt[4]{{(1 - v^{2} /c^2 )^{2} + 4a^{2} }}$,  (see
designations $a$ in \cite{2}-\cite{4}) the Lorentz's transformation also
may be used. In that case for the moving electrical charged particle
possessing velocity $v\simeq\infty$ and energy $E=\sqrt{2}E_{0}$, near to
the device playing role of the detector, there are following effects can
be probably used for detection of the fact of transit of a particle:\\
 a) In the real physical world any of the physical quantities can not be equal
infinity, so we shall introduce for designation of maximal velocity of a
particle designation $v_{m}$ ($v_{m}$ is the velocity of a faster than
light particle for which the energy loss accompanying by increase of
velocity is compensated by magnification of the energy gained from a
medium in which the beam of the particles flow by, i.e. the velocity of a
particle becomes practically stationary value, for example be in
thermodynamic equilibrium with the relic radiation that gives the particle
the velocity $v_{m}$$\sim500c$ ). There are an  almost instantaneous
impulses of electrical and magnetic fields from an electrical current
formed by transit of the faster than light particle through the media.
These impulses could be discovered by detectors that are capable to detect
super short impulses electrical or magnetic fields;\\ b)The kinetic energy
of a faster than light particle at $v>c$ looks like $E_{k} \approx
\sqrt{2}E_{0} c^{2}/v_{m}^{2}$. The transfer of parts of this energy
basically can be registered by high precision detector's (counters of
prompt particles for example based on use of an inner photoelectric
effect) in case when a faster than light particle has the collision at a
proton, nuclear or electron;\\
 c) In the lengthy detector filled by substances with large density
(small free length of collisions  for particles) will arise the multiple
collisions of faster than light particles with atoms. It can gives energy
transition from substances to a  faster than light particle and by that to
decreasing of its velocity. The power transmission from medium to a
particle will gives in decreasing of  temperature of medium and besides
gives the radiation of Cherenkov-Vavilov type (in an region of frequencies
defined by number of collisions with atoms of the substance of the
detector);\\
 d) When the faster than light particle fly through the substance with many
energy levels with negative temperatures as result may be lost of energy
of substance without radiation  and decreasing of negative temperature of
active optical substance. Physical laws do not forbid all numbered methods
of detection for ordinary particles with faster than light velocities and
their experimental realization (as well as many other method's are based
on an energy exchange of a faster than light particle with energy of
medium) are possible. The numbered methods realization depends on the
value of maximal velocity $v_{m}$ .

\section{Are the particles with $v>c$ and real mass exist in nature?}

Let us put the question: are the faster than light particles exist in our
world? When and where such particles can be discovered?  As the one of
consequences of the theory of fractal time (see \cite{1}), the particles
with velocities exceeding  the velocity of light must have  an energy
exceeding their the  rest energy $E_{o}$ in $10^{3}$ times. Such particles
may be borne for example  by explosions of stars ( in that case it is
possible to expect the appearance of the maximum in the spectrum of
${\gamma}$- quanta for the energies $E_{0}10^{3}$ or at the first moments
of "big bang" when temperatures of the early Universe exceed $10^{16}K$.
If a neutrinos have the rest mass and its rest energy are small and have
the order (or less) $1ev.$ the neutrinos with faster then light velocities
may be produced  by stars, by nuclei explosions and in the reactions of
thermonuclear controlled syntheses. May some super civilization use the
faster then light particles, if this civilization has the technology of
receiving the beams of such particle, for record and translating
information with the faster than light velocities ? In that case it is
necessary  to seek such particles by mentioned above (or similar) methods.

\section{Conclusion}

On the basis of the above-stated treatment of possibilities of detection
of  the particles with the faster than light speed, it is possible to make
a deduction: the prohibitions for transfer and receiving of the
information with faster than light  speed are absent (if the theory
\cite{1}-\cite{4} are valid).  The question about  an existence of the
ordinary  particles (protons, electrons, neutrinos) with velocities faster
than light and the real mass in nature ( that question was presented (and
decided) for the first time in the paper \cite{2} as one of the
consequences of the theory of almost inertial systems that lays beyond of
the special relativity and coincides with SR in the case of ideal inertial
systems) is now unsolved. The search of taxions continue more than thirty
years. I don't mention about the optical taxions. The existence of the
optical taxions do not contradict the SR and apparently they are
discovered. I think that only careful experimental search of the ordinary
particles  with the real mass (the faster than light particles) and
experiments that may examine the fractal theory of time \cite{1} may throw
light on this very interesting problem.\\\\ We suggest to carry out the
experiments for receiving by accelerating the protons with energies equal
${\sim{10^{12}ev.}}$  (that gives a protons the velocity equal the speed
of light if the theory \cite{1}-\cite{2} are valid) , then to verify the
predictions of the theory that presented in this paper and papers
\cite{2}-\cite{4}

\end{document}